\documentclass[prc,twocolumn,showpacs,amsmath,amssymb]{revtex4}

\usepackage{times}
\usepackage{graphicx}
\usepackage{dcolumn}
\usepackage{bm}
\usepackage{amstext}

\begin{document}

\title{The Oxygen core inside the Magnesium isotopes}
\author{M. Bhuyan}

\affiliation{Institute of Physics, Sachivalaya Marg, 
Bhubaneswar-751 005, India.}

\email{bunuphy@iopb.res.in}
\date{\today}

\begin{abstract}
We have studied the ground state bulk properties of magnesium isotopes
using axially symmetric relativistic mean field formalism. The BCS pairing
approach is employed to take care of the pairing correlation for the open
shell nuclei. The contour plot of the nucleons distribution are analyzed
at various parts of the nucleus, where clusters are located. The presence
of an $^{16}$O core along bubble like $\alpha$-particle(s) and few
{\it nucleons} are found in the Mg isotopes.
\end{abstract}


\maketitle

\section{Introduction}
The internal configuration of a nuclear system plays an important role
on the stability, which is connected to the cluster radioactivity. In
1984, the cluster radioactivity was discovered \cite{0} and the consequent
excitement led to several experiments at different places all over the world.
Finally, {\it Milano} had developed the experimental technique to investigate
such extremely rare mode of decay \cite{1,1a}. Further, the crucial
experiments \cite{1,2,3}are most notable inputs towards the description of
cluster radioactivity. The review of last three decades show a regular
interest covering both theory and experiment to explain the aspects of cluster
and its exotic radioactivity till date \cite{4}.

After the confirmed identification of nuclear sub-structure (clustering),
few questions arises in mind: (i) are they initially present inside the
parent nucleus (ii) how they look like and (iii) what are the constituent
of these clusters. Hence, it is important to see the preformed cluster(s)
inside the decaying nucleus. Several techniques are existed in literature
to understand the clustering structure of the nuclear system. The prediction
of $3-\alpha$ structure of $^{12}$C and $4-\alpha$ of $^{16}$O are already
observed experimentally \cite{5,6,7,8,9,10}. The presence of these clusters
inside a nucleus are due to the random distribution of density. Statistically,
the large magnitude in density distribution at cluster region from its
surroundings indicates the maximum population of the nucleon(s). This may be
a reason for cluster(s) decay. The formation probability and decay half-life
of these nuclei have been studied from last three decades by different peoples
over the world \cite{11,12,13,14,15}.
From these studies, one can find the most possible clusters are $^4$He,
$^8$Be, $^{12}$C, $^{16}$O, $^{20}$Ne, $^{24}$Mg and $^{28}$Si having $N=Z$,
which are integral multiple of the $\alpha$-cluster ($n\cdot \alpha$). Hence
one can say, in case of lighter mass regions, the $\alpha$-particle could be
the constituent of emitted nuclei.  Recently, the theoretical predictions of
clusters in lighter mass nuclei \cite{add1,add1a,16,17,18} and some interesting
phenomenon such as low-energy spectra of Mg-isotopes
\cite{yao11,bend03,ring06,ring06a,eunja13}, motivate us to look the the internal 
configuration of these nuclei using relativistic mean field formalism. In
this context, the present work directed to a particular case for Mg isotopes
is taken up to examine the preformed clusters and their constituents. It is
worth mentioning that, the RMF theory is very much successful in explaining
the sub-atomic nuclei \cite{19,20,21} and the decays of these nuclei
\cite{22,23}. Because of its applicability, we have used RMF theory with
the recently developed {\it NL3$^*$} \cite{23a} and {\it NL075} \cite{add2}
parameter sets to study the clustering phenomena.

The paper is organized as follows: The relativistic mean field theory is
described briefly in Sec. 2 including BCS pairing approach. In Sec. 3, the
details of our calculation and results are discussed. Finally, the summary
and concluding remarks are given in Section 4.

\section{The relativistic mean-field (RMF) method}
In last few decades, the RMF theory is applied successfully to study
the structural properties of nuclei throughout the periodic table
\cite{24,25,26,27,28,29} starting from proton to neutron drip-lines.
The relativistic Lagrangian density for nucleon-meson many-body system
is expressed as \cite{27,29},
\begin{eqnarray}
{\cal L}&=&\overline{\psi_{i}}\{i\gamma^{\mu}
\partial_{\mu}-M\}\psi_{i}
+{\frac12}\partial^{\mu}\sigma\partial_{\mu}\sigma
-{\frac12}m_{\sigma}^{2}\sigma^{2}\nonumber\\
&& -{\frac13}g_{2}\sigma^{3} -{\frac14}g_{3}\sigma^{4}
-g_{s}\overline{\psi_{i}}\psi_{i}\sigma-{\frac14}\Omega^{\mu\nu}
\Omega_{\mu\nu}\nonumber\\
&&+{\frac12}m_{w}^{2}V^{\mu}V_{\mu}
+{\frac14}c_{3}(V_{\mu}V^{\mu})^{2} -g_{w}\overline\psi_{i}
\gamma^{\mu}\psi_{i}
V_{\mu}\nonumber\\
&&-{\frac14}\vec{B}^{\mu\nu}.\vec{B}_{\mu\nu}+{\frac12}m_{\rho}^{2}{\vec
R^{\mu}} .{\vec{R}_{\mu}}
-g_{\rho}\overline\psi_{i}\gamma^{\mu}\vec{\tau}\psi_{i}.\vec
{R^{\mu}}\nonumber\\
&&-{\frac14}F^{\mu\nu}F_{\mu\nu}-e\overline\psi_{i}
\gamma^{\mu}\frac{\left(1-\tau_{3i}\right)}{2}\psi_{i}A_{\mu} .
\end{eqnarray}
Here $M$, $m_{\sigma}$, $m_{\omega}$ and $m_{\rho}$ are the masses
for nucleon, ${\sigma}$, ${\omega}$ and ${\rho}$- mesons and $\psi$
is its Dirac spinor. The field for the ${\sigma}$-meson is denoted
by ${\sigma}$, for ${\omega}$-meson by $V_{\mu}$ and for ${\rho}$-meson
by $R_{\mu}$. The coupling constants for $\sigma-$, $\omega$ and
$\rho-$ mesons are $g_s$, $g_{\omega}$ and $g_{\rho}$, respectively.
The self-coupling constants, $g_2$ and $g_3$ are for $\sigma-$meson
and $e^2/4\pi$=1/137 is the fine structure constant for photon. From
the above Lagrangian we obtained the field equations for the nucleons
and mesons. These equations are solved by expanding the upper and lower
components of the Dirac spinors and the boson fields in an axially deformed
harmonic oscillator basis with an initial deformation $\beta_{0}$. The set
of non-linear coupled equations is solved numerically by a self-consistent
iteration method \cite{26,29b,29c,29d}. The center-of-mass motion
energy correction is estimated by the usual harmonic oscillator formula
$E_{c.m.}=\frac{3}{4}(41A^{-1/3})$. The quadrupole deformation parameter
$\beta_2$ is evaluated from the resulting proton and neutron quadrupole
moments, as $Q=Q_n+Q_p=\sqrt{\frac{16\pi}5} (\frac3{4\pi} AR^2\beta_2)$.
The root mean square (rms) matter radius is defined as
$\langle r_m^2\rangle={1\over{A}}\int\rho(r_{\perp},z) r^2d\tau$, where
$A$ is the mass number, and $\rho(r_{\perp},z)$ is the deformed density.
The total binding energy and other observables are also obtained by using
the standard relations, given in \cite{29a}. To deal with the open shell
nuclei, we have adopted BCS-pairing method in a constant gap approximation
\cite{31,32}. The expression for the pairing energy is given by:
\begin{equation}
E_{pair}=-G\left[\sum_{i>0}u_{i}v_{i}\right]^2,
\end{equation}
where {\it G} is the pairing force constant and $v_i^2$ and
$u_i^2 =(1 - v_i^2)$ are the occupation probabilities. The variational
procedure with respect to the occupation numbers $v_i^2$, gives the BCS
equation;
\begin{equation}
2\epsilon_iu_iv_i-\triangle(u_i^2-v_i^2)=0,
\end{equation}
and the gap $\Delta$ is defined by
\begin{equation}\triangle=G\sum_{i>0}u_{i}v_{i}.
\end{equation}
The occupation number $n_i$ is defined as
\begin{equation}
n_i=v_i^2=\frac{1}{2}\left[1-\frac{\epsilon_i-\lambda}
{\sqrt{(\epsilon_i-\lambda)^2+\triangle^2}}\right].
\end{equation}
The constant gap pairing for protons and neutrons are
taken from Refs. \cite{add3}:
\begin{eqnarray}
\triangle_p &=& RB_s e^{sI-tI^2}/Z^{1/3}\;\;\;\;\;\mbox{and}\\
\triangle_n &=& RB_s e^{-sI-tI^2}/A^{1/3},
\end{eqnarray}
with $R$ = 5.72, $s$ = 0.118, $t$ = 8.12, $B_s$ = 1, and
$I=(N-Z)/(N+Z)$.
The chemical potentials $\lambda_n$ and $\lambda_p$ are determined
by the particle numbers for protons and neutrons and the pairing
energy is given as
\begin{equation}
E_{pair}= -\frac{\Delta^2}{G} =-\triangle\sum_{i>0}u_{i}v_{i}.
\end{equation}
From the Eqn. (8), it is clear that the pairing energy $E_{pair}$
is not static even in the constant gap approach because it depends
on the occupation probabilities $u_i^2$ and $v_i^2$, which is directly
connected with the deformation parameter $\beta_2$ (spherical or
deformed) near the Fermi surface. It is well known that
$E_{pair}$ diverges if it is extended to an infinite configuration
space for a constant gap $\Delta$ and strength parameters $G$. However,
a constant pairing gap is taken for simplicity of the calculations.
Within this pairing approach, it is shown that the results for binding
energies and quadrupole deformations are almost identical with the
predictions of Relativistic-Hartree–Bogoliubov (RHB) approach
\cite{34,35,36,37,38,39} near the proton-drip line.
\begin{table*}
\caption{The RMF (NL3$^*$ and NL075) results for binding energy, the
quadrupole deformation parameter $\beta_2$, and charge radii for $Mg$-
isotopes are compared with the experimental data$^{47,48,49}$. The
energy in $MeV$ and radius in $fm$.
}
\renewcommand{\tabcolsep}{0.12cm}
\renewcommand{\arraystretch}{1.4}
\begin{tabular}{cccccccccc}
\hline
\hline
Nucleus&\multicolumn{3}{c}{RMF (NL3$^*$)}& \multicolumn{3}{c}{RMF (NL075)}
& \multicolumn{3}{c}{Experiment} \\
\hline
& BE & $\beta_{2}$ & $r_{ch}$ & BE & $\beta_{2}$ & $r_{ch}$ & BE & $\beta_{2}$ & $r_{ch}$ \\
\hline
$^{20}$Mg & 135.1 & 0.02 & 3.169 & 133.3 & 0.04 & 3.250 & 134.5 (0.028) &              & \\
$^{22}$Mg & 166.2 & 0.39 & 3.159 & 164.7 & 0.46 & 3.241 & 168.5 (0.013) & 0.58 (0.011) & \\
$^{24}$Mg & 193.5 & 0.45 & 3.125 & 190.1 & 0.52 & 3.210 & 198.2 (0.0002)& 0.60 (0.008) & 3.057 (0.002) \\
$^{26}$Mg & 212.5 & 0.33 & 3.061 & 209.9 & 0.51 & 3.185 & 215.1 (0.0003)& 0.48 (0.010) & 3.033 (0.002) \\
$^{28}$Mg & 228.1 & 0.26 & 3.061 & 224.7 & 0.40 & 3.159 & 231.6 (0.002) & 0.49 (0.035) & \\
$^{30}$Mg & 240.5 & 0.56 & 3.072 & 236.8 & 0.23 & 3.110 & 241.1 (0.008) & 0.43 (0.019) & \\
$^{32}$Mg & 250.5 & 0.00 & 3.091 & 249.2 & 0.01 & 3.094 & 249.7 (0.013) & 0.47 (0.043) & \\
$^{34}$Mg & 256.5 & 0.32 & 3.149 & 252.9 & 0.33 & 3.159 & 256.5 (0.238) &              & \\
\hline
\hline
\end{tabular}
\end{table*}

\section{Method of Calculations and Discussions}
We have carried out the numerical calculations using maximum oscillator
quanta $N_F=N_B=10$ for Fermion and boson. To test the convergence of
the solutions, few calculations are done with $N_F$ = $N_B$ = 12 also.
The variation of these two solutions are $\leq 0.02\%$ on binding energy
and 0.01$\%$ on nuclear radii for drip-line nuclei. This implies that
the used model space is good enough for the considered nuclei. The number
of mesh points for Gauss-Hermite and Gauss-Lagurre integration are $20$
and $24$, respectively. For a given nucleus, the maximum binding energy
corresponds to the ground state and other solutions are obtained at various
excited intrinsic states. In our calculations, we obtained different
nucleonic potentials, densities, single-particle energy levels,
root-mean-square (rms) radii, deformations and binding energies. These
observables explain the structure and sub-structure for a nucleus in
a given state.

\subsection{The Binding Energy and Charge Radius}
The nuclear bulk properties are mostly responsible for the internal
configuration (arrangement) of the nuclei. In this context, we have
calculated the bulk properties like binding energy (BE), nuclear
charge radii ($r_{ch}$) and the quadrupole deformation parameter
($\beta_2$) for the ground state of $^{20-34}$Mg. The results obtained
from NL3* and NL075 are compared with the experimental data \cite{40,41,41a}
in Table 1. The ground state solution for all the isotopes are followed by
a deformed prolate configuration which are comparable with the experimental
data \cite{41a}. Further, a careful inspection reveal that the prediction
of $NL3^*$ for binding energy and radii is slightly better with experimental
observation. However, the quadrupole deformation is more closer to the
experiment in case of $NL075$.  In general, both the forces predict similar
results which are quite close to the experimental values.

\begin{figure}
\includegraphics[width=1.0\columnwidth]{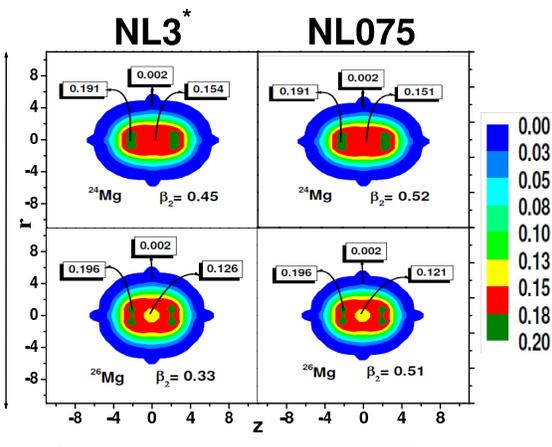}
\vspace{0.6cm}
\caption{The internal sub-structure of $^{24,26}$Mg in the ground state
configuration.
}
\end{figure}

\begin{figure}
\vspace{0.5cm}
\includegraphics[width=1.0\columnwidth]{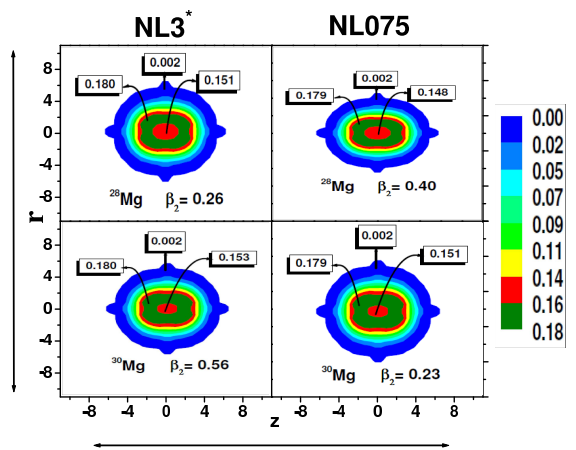}
\vspace{0.6cm}
\caption{The internal sub-structure of $^{28,30}$Mg in the ground state
configuration.
}
\end{figure}

\subsection{The Clustering and Sub-atomic Nuclei}
The internal structure of a nucleus depends on the density distributions
of the proton, neutron and matter for a given state. Here, the densities
are obtained from RMF (NL3* and NL075) in the positive quadrant of the
plane parallel to the symmetry $z-$axis. These are evaluated in the
$\rho z$ plane, where $\rho=x=y=r_{\bot}$. It is to be noted that, both
the axes $z$ and $\rho$ are conserved in our formalism under the space
reflection symmetry. Now we can obtain the complete picture of a nucleus
in the $\rho z $ plane by reflecting the first quadrant to other quadrants.
The contour plotting of density along with the color code for the ground
state of $^{24-26}$Mg and $^{28-30}$Mg are shown in Figs. 1 and 2,
respectively. In the isotopic series, the ground states belong to deformed
prolate solution (see Table 1), which is also reflected in the figures. From
the color code, one can identify the clustering structures in $Mg$ nuclei.
For example, the color code with deep green corresponds to maximum density
$\rho \sim$ 0.18 $fm^{-3}$ and the deep blue bearing the minimum value of
$\rho \sim$ 0.001 $fm^{-3}$. (In black and white figures, the color code
is read as deep black with maximum density to outer gray as minimum density
distribution). A careful inspection of the figures show the formation of
various clusters inside the nuclei. This region has a very high probability
of preformation and decaying from its interior. The constituents of this
cluster depends on the size and the magnitude of the density of the region.
To determine, these sub-nuclear structure (cluster) inside the nucleus, it
is important to know the volume of the cluster, i.e., ranges or area covered
by the cluster.

\begin{table*}
\caption{The RMF (NL3*) results for cluster (s) inside the nuclei and the
corresponding states along with the range of the cluster and the respective
densities are listed.}
\renewcommand{\tabcolsep}{0.18cm}
\renewcommand{\arraystretch}{1.2}
\begin{tabular}{ccccccccccccc}
\hline
\hline
Nucleus& $\beta_2$ & Range ($r_1$,$r_2$; $z_1$,$z_2$) & $Z_{clus.}$
& $N_{clus.}$ & Cluster & Diffused nucleons \\
\hline
RMF (NL3$^*$) \\
$^{24}$Mg & 0.45 & -1.09,1.09; 2.76,2.82  & 1.98 & 2.01 & $^{4}$He & $8n$ \& $8p$\\
          &       & -1.09,1.09; -1.76,-2.82& 1.98 & 2.01 & $^{4}$He &              \\
RMF (NL075) \\
$^{24}$Mg & 0.52 & -1.09,1.09; 2.76,2.82  & 1.98 & 2.01 & $^{4}$He & $8n$ \& $8p$\\
          &       & -1.09,1.09; -1.76,-2.82& 1.98 & 2.01 & $^{4}$He &              \\
\hline
RMF (NL3$^*$) \\
$^{26}$Mg & 0.33 & -1.1,1.1; 1.68,2.35    & 1.92 & 2.03 & $^{4}$He & $10n$ \& $8p$ \\
          &       & -1.1,1.1;-1.68,-2.35   & 1.93 & 2.03 & $^{4}$He & \\
RMF (NL075) \\
$^{26}$Mg & 0.51 & -1.1,1.1; 1.68,2.35    & 1.92 & 2.03 & $^{4}$He & $10n$ \& $8p$ \\
          &       & -1.1,1.1;-1.68,-2.35   & 1.93 & 2.03 & $^{4}$He & \\
\hline
RMF (NL3$^*$) \\
$^{28}$Mg & 0.26 & -1.84,1.84; -0.81,0.81 & 0.06 & 3.98 & $4n$     & $4n$ \& $4p$ \\
          &       &  -1.84,1.84; 2.67,2.67 & 7.89 & 8.08 & $^{16}$O & \\
RMF (NL075) \\
$^{28}$Mg & 0.40 & -1.84,1.84; -0.81,0.81 & 0.06 & 3.98 & $4n$     & $4n$ \& $4p$ \\
          &       &  -1.84,1.84; 2.67,2.67 & 7.89 & 8.08 & $^{16}$O & \\
\hline
RMF (NL3$^*$) \\
$^{30}$Mg & 0.56 & -1.99,1.99; 0.81,0.81  & 0.04 & 5.98 & $6n$     & $4n$ \& $4p$ \\
          &       & -1.83,1.83; 2.62,2.62  & 7.89 & 8.12 & $^{16}$O & \\
RMF (NL075) \\
$^{30}$Mg & 0.23 & -1.99,1.99; 0.81,0.81  & 0.04 & 5.98 & $6n$     & $4n$ \& $4p$ \\
          &       & -1.83,1.83; 2.62,2.62  & 7.89 & 8.12 & $^{16}$O & \\
\hline
\hline
\end{tabular}
\label{Table 1}
\end{table*}

The dimension of the cluster is estimated from the contour plot, defines the
lower and upper limit of the integral in equation (9) [$r_{\bot}$
($r_1$, $r_2$) and $z$ ($z_1$,$z_2$)]. It is worth mentioning that the ranges
are fixed by graphical method which is guided through eyes and may result some
uncertainty. The values of the ranges for different clusters for some of the
Mg isotopes are listed in Table 2. The formula used to identify the ingredient
of the cluster is given by \cite{19,21}:
\begin{equation}
n=\int_{z_1}^{z_2}\int_{r_1}^{r_2}\rho (z,r_{\bot})d\tau,
\end{equation}
where, $n$ is the number of neutrons $N$ or protons $Z$ or mass $A$ and
$z$ ($z_1$, $z_2$), $r_{\bot}$ ($r_1$ , $r_2$) are the ranges. From the
estimated proton and neutron numbers, we determine the mass of the cluster
inside the nucleus. The obtained clusters for ground state of Mg isotopes
are listed in Table 2. From the table, we noticed the presence of $^{16}$O
and two $^4$He along with few neutrons in $^{24-30}$Mg.

As a further confirmatory test, here we compare the central density
of $^{16}$O and $^4$He to the cluster density inside the core of
$Mg$-isotopes. Normally, the magnitude of the central density of
$^4$He and $^{16}$O are $\sim 0.20$ and $0.16$, respectively. Here,
the maximum value of the density correspond to these cluster region
within the range $\sim 0.14-0.17$, which matches to the normal density
of $^{16}$O. Hence, this confirms, the constituent nucleus inside the
cluster (s) region may either $^{16}$O or the condensate
$4 \cdot \alpha$-particles. It could accept that the present study is
a quantitative analysis for the internal structure of $^{24}$Mg. This
finding of $^{16}$O and $^{4}$He bubbles in $^{24}$Mg is irrespective
of NL3* and NL075 forces, which shows the universality of the RMF
formalism. As these are preformed clusters inside the nuclei, they may
have high decaying probability. The existence of such bubble inside Mg
could be an interesting experimental observation.

\section{SUMMARY AND CONCLUSIONS}
Concluding, we have presented the gross nuclear properties like binding
energy, deformation parameter $\beta_2$, charge radii $r_{ch}$ and the
nucleonic density distributions for the isotopic chain $^{20-34}$Mg using
an axially deformed relativistic mean field formalism with $NL3^*$ and
$NL075$ parameter sets. The results of our calculations show qualitative
and quantitative agreement with the experimental observations. We found
deformed prolate ground states solution for Mg isotopes, which are
consistent with the experimental data. Analyzing the nuclear density
distributions, the internal structure i.e. the clusters of Mg isotopes
are identified. We found sub-structure like $^{16}$O or $4 \cdot \alpha$
condensate types along with few more neutrons inside Mg isotopes. It is
also noticed that the cluster structure of a nucleus remain unaffected
for different force parameters as long as the solution for that nucleus
exist. It is interesting to see these evaporation residues like $^{16}$O
and $^{4}$He for Mg-isotopes in laboratory.

\section*{Acknowledgments}
This work is supported in part by Council of Scientific \& Industrial
Research (File No.09/153 (0070) /2012-EMR-I).

\end{document}